\documentclass[aps,pra,superscriptaddress,floatfix,showpacs,10pt,twocolumn]{revtex4}
\usepackage[dvips]{graphicx}
\usepackage{amsmath}
\begin{document}
\title{Entanglement purification for arbitrary unknown ionic states via linear optics}
\author{Ming Yang}
\email{mingyang@ahu.edu.cn}
\author{Wei Song}
\author{Zhuo-Liang Cao}
\email{zlcao@ahu.edu.cn} \affiliation{Anhui Key Laboratory of
Information Material \& Devices, School of Physics \& Material
Science,Anhui University, Hefei, 230039, People's Republic of
China}
\begin{abstract}
An entanglement purification scheme for arbitrary unknown(mixed
and pure non-maximally) entangled ionic states is proposed by
using linear optical elements. The main advantage of the scheme is
that not only two-ion maximally entangled pairs but also four-ion
maximally entangled pairs can be extracted from the less entangled
pairs. The scheme is within current technology.
\end{abstract}
\pacs{03.67.Hk, 03.67.Mn, 03.67.Pp, 42.50.Dv} \maketitle

\section{INTRODUCTION}
Entanglement~\cite{en} first presented by Schr\"{o}dinger is a
critical manifestation of quantum mechanics. Resulting from its
non-locality property, entanglement has become more and more
important resource in Quantum Information Processing(QIP). All of
the applications~\cite{tele, cryp, comput, hidden} of entanglement
work perfectly only with the pure maximally entangled states.
Because quantum entanglement can only be produced
locally~\cite{mix}, the entangled objects must be distributed
among distant users for Quantum Communication purpose. Due to the
impossibility that one quantum system can be isolated from the
environment absolutely, the entanglement of the entangled objects
will decrease exponentially with the propagating distance of the
objects, and the practically available quantum entangled states
are all non-maximally entangled states or the more general
case--mixed states. So, if nothing has been done on the
distributed states before used in Quantum Communication, the long
distance Quantum Communication~\cite{lmd} is impossible. To
overcome the dissipation and decoherence, various schemes of
entanglement distillation~\cite{distillation, distillation1},
entanglement concentration~\cite{concentation, concentration1} and
entanglement purification~\cite{puri, jwp, jwp1, simon,
linden,horodecki, gisin, thew, unknown,wdur, lmd1, sjvan, remero}
have been proposed. Alternatively, Quantum
Repeater~\cite{repeater} also can be used to overcome this
difficulty. The main processes of a Quantum Repeater are composed
of entanglement purification~\cite{puri} and entanglement
swapping~\cite{concentration1}, and the main task of it is still
to realize entanglement purification. So we will mainly discuss
the entanglement purification process. Entanglement purification
is a method that can extract a small number of entangled pairs
with relatively high degree of entanglement from a large number of
less entangled pairs using only local operations and classical
communication. In the original entanglement purification
scheme~\cite{puri}, C-NOT operations construct the main step of
the purification process. But, in experiment, there is no
implementation of C-NOT operations can meet the error rate level,
which is needed for the logic gates in long distance Quantum
Communication~\cite{repeater}.So more and more attention are
focused on finding the realizable schemes for entanglement
purification. J.-W. Pan \emph{et al} use the Polarization Beam
Splitter(PBS)~\cite{jwp} to replace the C-NOT gate needed in the
original scheme~\cite{puri}, and can get the newly-obtained
polarization-entangled photon pairs with a larger fraction of
fidelity. Most of the above entanglement purification schemes are
theoretical ones. Recently, significant progresses on entanglement
purification have been achieved in experiment~\cite{jwp1,
distillation}. P.G. Kwiat \emph{et al} proposed a experimental
entanglement distillation scheme for pure non-maximally and mixed
polarization-entangled photon states using partial
polarizers~\cite{distillation}. Following the theoretical
proposal~\cite{jwp}, J.-W. Pan \emph{et al} successfully realize
the entanglement purification of general mixed states of
polarization-entangled photon pairs using linear optics elements
in experiment~\cite{jwp1}.

From the previous entanglement purification schemes, we conclude
that most of them can only apply to the polarization-entangled
photon pairs. There is few schemes for distillation~\cite{me, me1}
and purification of atomic and ionic entangled states in the
literature. Although, photons are the attractive carriers of
information for the implementation of Quantum Communication, ions
are also the preferred carrier for quantum information, because
the realization of Quantum Computer and Quantum Computation relies
on the optimal quantum carriers, which should can be integrated.
So the purification of ionic entangled states is of practical
significance not only in Quantum Communication but also in Quantum
Computation.

Inspired by J.-W, Pan's proposal~\cite{jwp} for entanglement
purification and X.-X. Zhou's proposal for Non-distortion Quantum
Interrogation(NQI)~\cite{nqi}, we will propose, in this paper, an
entanglement purification scheme for arbitrary
unknown~\cite{unknown, jwp1} mixed entangled ionic states by using
Beam Splitters(BS) and polarization-sensitive single photon
detectors(D). For the arbitrary unknown non-maximally entangled
pure states, it also works.Through analysis, we can get a
near-perfect maximally entangled ionic states from the mixed
entangled ionic states, provided we repeat the scheme several
times. From the pure non-maximally entangled states, we can get
the perfect maximally entangled ionic states probabilistically. We
can decide whether the purification procedure succeeds by
operating single photon measurement on each side.

\section{ENTANGLEMENT PURIFICATION FOR MIXED STATES}
\begin{figure}
\includegraphics[scale=0.27, angle=270]{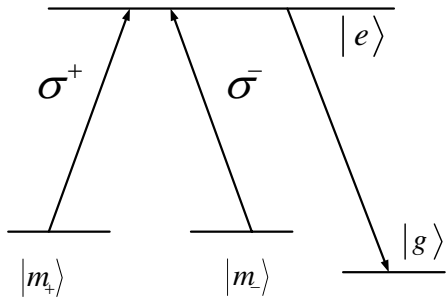}
\caption{\label{level}Level configuration of the ions used in the
scheme. The ions, which are in the degenerate states
$|m_{+}\rangle\textrm{ and }|m_{-}\rangle$, can be excited into
the unstable excited state $|e\rangle$ by absorbing one
$\sigma^{+}\textrm{ or }\sigma^{-}$ polarized photon, then it can
decay to the stable ground state $|g\rangle$ with a scattered
photon rapidly.}
\end{figure}

For communication purpose, the two distant users Alice and Bob
should share maximally entangled states:
\begin{subequations}
\begin{equation}
|\Phi^{+}\rangle_{12}=
\frac{1}{\sqrt{2}}(|m_{+}\rangle_{1}|m_{+}\rangle_{2}+|m_{-}\rangle_{1}|m_{-}\rangle_{2}),
\end{equation}
\begin{equation}
|\Psi^{+}\rangle_{12}=
\frac{1}{\sqrt{2}}(|m_{+}\rangle_{1}|m_{-}\rangle_{2}+|m_{-}\rangle_{1}|m_{+}\rangle_{2}).
\end{equation}
\end{subequations}
These are two Bell states for two ions. One of the two ions is at
Alice' side, the other at Bob's. Here, $|m_{+}\rangle\textrm{ and
}|m_{-}\rangle$ are two degenerate metastable states of ions. The
ions can be excited from $|m_{+}\rangle\textrm{ or }|m_{-}\rangle$
to the excited states $|e\rangle$ by absorbing one
$\sigma^{+}\textrm{ or }\sigma^{-}$ circular polarization photon
with unit efficiency. The excited state $|e\rangle$ is not a
stable one, so the ions in that state will decay rapidly to the
stable ground state $|g\rangle$ with a scattered photon
$|S\rangle$. This process can be expressed as:
\begin{equation}
a_{\pm}^{+}|0\rangle|m_{\pm}\rangle\longrightarrow|S\rangle|g\rangle.
\end{equation}
The level configuration of the ions is depicted in
Fig.\ref{level}.
But, for communication purpose, the two ions must be distributed
to different locations. During the transmission process,
entanglement will inevitably degrade. So the entangled states
after distribution are usually  mixed ones. Suppose that the mixed
state to be purified is in the form:
\begin{equation}\label{old}
\rho_{AB}=F|\Phi^{+}\rangle_{AB}\langle\Phi^{+}|+(1-F)|\Psi^{+}\rangle_{AB}\langle\Psi^{+}|.
\end{equation}
Because a general mixed state can be rotated into the form in
equation (\ref{old}), the discussion on the state in equation
(\ref{old}) applies to general mixed cases~\cite{jwp}. Further, to
complete the purification scheme, we suppose that Alice and Bob
have shared an ionic ensemble, each pair of which can be described
by the state in equation (\ref{old}). Here,
$F=\langle{\Phi^{+}}|\rho_{AB}|{\Phi^{+}}\rangle$ is the fidelity
of the pairs with respect to $|{\Phi^{+}}\rangle$.

Next, we will discuss the purification procedure in details. To
complete the purification process, we must carry out operations on
two pairs of the ensemble. We denote the four ions of the two
pairs as $1, 2 \textrm{ and } 3, 4$, and the total state of the
two pairs before purification can be regarded as a probabilistic
mixture of four pure
states:$|\Phi^{+}\rangle_{12}|\Phi^{+}\rangle_{34}$ with
probability $F^{2}$, $|\Phi^{+}\rangle_{12}|\Psi^{+}\rangle_{34}$
with probability $F(1-F)$,
$|\Psi^{+}\rangle_{12}|\Phi^{+}\rangle_{34}$ with probability
$(1-F)F$, and $|\Psi^{+}\rangle_{12}|\Psi^{+}\rangle_{34}$ with
probability $(1-F)^{2}$.

\begin{figure}
\includegraphics[scale=0.27, angle=270]{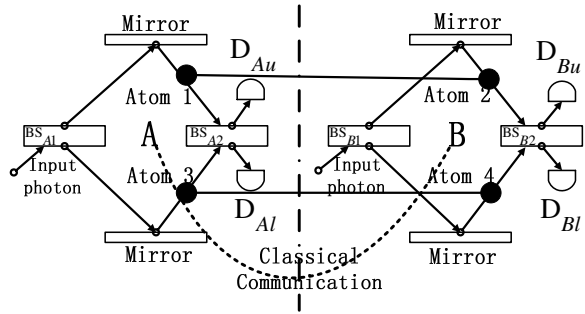}
\caption{\label{setup}The setup for purification scheme. Alice
places two ions 1,3 which are at her side on the two arms of
$M_{A}$, ion 1 on upper path and ion 3 on lower one by using the
trapping technology~\cite{trapping}, and analogously for the two
ions 2, 4 at Bob' side. One $\sigma^{+}$ polarized photon at each
side will be superimposed on the first two BS($BS_{A1}\textrm{ and
}BS_{B1}$) of $M_{A}\textrm{ and }M_{B}$ respectively. After the
first BS the photon will take two possible pathes($u$ denotes the
upper path and $l$ denotes the lower one). Reflected by two
mirrors, the two possible pathes will re-combined at the second BS
($BS_{A2}\textrm{ and }BS_{B2}$). Because the two ions at one side
are initially placed on the two optical pathes, the ions and the
photon will interact. This interaction will generate a shift of
the interference after the second BS($BS_{A2}\textrm{ and
}BS_{B2}$).Then through single photon measurement after the two
second BS($BS_{A2}, BS_{B2}$), Alice and Bob can compare their
measurement results via classical communication. If the two lower
output ports($D_{Al}, D_{Bl}$) all fire, the purification
succeeds.}
\end{figure}

The main setup, depicted in Fig.\ref{setup}, are two Mach-Zehnder
interferometers($M_{A}\textrm{ and }M_{B}$) located at Alice and
Bob's side respectively.

We suppose the input photon at Alice side is $\sigma^{+}$
polarized, and it is superimposed on $BS_{A1}$ at the left lower
input port of $M_{A}$, and analogously for the description of
Bob's side. The effect of the BS on the input photon can be
expressed as:
\begin{subequations}
\begin{equation}
a_{l,\pm}^{+}|0\rangle_{i}\stackrel{\textrm{BS}}{\longrightarrow}
\frac{1}{\sqrt{2}}(a_{u,\pm}^{+}+ia_{l,\pm}^{+})|0\rangle_{i},
\end{equation}
\begin{equation}
a_{u,\pm}^{+}|0\rangle_{i}\stackrel{\textrm{BS}}{\longrightarrow}
\frac{1}{\sqrt{2}}(a_{l,\pm}^{+}+ia_{u,\pm}^{+})|0\rangle_{i}.
\end{equation}
\end{subequations}
where $l\textrm{ and }u$ denote optical pathes(lower and upper),
$i=A, B$, $a_{l,\pm}^{+}|0\rangle_{A}\textrm{ and }
a_{l,\pm}^{+}|0\rangle_{B}$ denote two input photons of
$M_{A}\textrm{ and } M_{B}$ respectively and $\pm$ denotes the
direction of polarization. The BS splits the wave function of the
input photon into two parts--reflected part and transparent one.
There will be a $\frac{\pi}{2}$ phase shift between the input
photon and the reflected wave function, and the transparent part
is synchronized with the input photon. The BS takes no effect on
the polarization of the input photon. These are critical to the
purification process.

To analyze the evolution of the total system, we will consider the
evolution of the following four product states of two ions. We
will consider the 1,3 ions case, and the result for ions 2, 4 are
same to that of ions 1, 3 case:
\begin{subequations}
\begin{eqnarray}
&a_{l,+}^{+}|0\rangle_{A}|m_{+}\rangle_{1}|m_{+}\rangle_{3}
\xrightarrow{BS_{A1}, Ions 1, 3, BS_{A2}}\nonumber\\
&\frac{1}{\sqrt{2}}
(|S\rangle_{1}|g\rangle_{1}|m_{+}\rangle_{3}+i|m_{+}\rangle_{1}|S\rangle_{3}|g\rangle_{3}),
\end{eqnarray}
\begin{eqnarray}
&a_{l,+}^{+}|0\rangle_{A}|m_{+}\rangle_{1}|m_{-}\rangle_{3}
\xrightarrow{BS_{A1}, Ions 1, 3, BS_{A2}}\nonumber\\
&\frac{1}{\sqrt{2}}
|S\rangle_{1}|g\rangle_{1}|m_{-}\rangle_{3}\nonumber\\
&+\frac{i}{2}(a_{u,+}^{+}+ia_{l,+}^{+})
|0\rangle_{A}|m_{+}\rangle_{1}|m_{-}\rangle_{3}
\end{eqnarray}
\begin{eqnarray}
&a_{l,+}^{+}|0\rangle_{A}|m_{-}\rangle_{1}|m_{+}\rangle_{3}
\xrightarrow{BS_{A1}, Ions 1, 3, BS_{A2}}\nonumber\\
&\frac{i}{\sqrt{2}}
|m_{-}\rangle_{1}|S\rangle_{3}|g\rangle_{3}+\frac{1}{2}(a_{l,+}^{+}+ia_{u,+}^{+}) |0\rangle_{A}\nonumber\\
&\times|m_{-}\rangle_{1}|m_{+}\rangle_{3},
\end{eqnarray}
\begin{eqnarray}
&a_{l,+}^{+}|0\rangle_{A}|m_{-}\rangle_{1}|m_{-}\rangle_{3}
\xrightarrow{BS_{A1}, Ions 1, 3, BS_{A2}}\nonumber\\
&ia_{u,+}^{+}|0\rangle_{A}|m_{-}\rangle_{1}|m_{-}\rangle_{3}.
\end{eqnarray}
\end{subequations}
Then we can give the evolution of the four probabilistic pure
states:
\begin{subequations}
\begin{eqnarray}
&F^{2}:\nonumber\\
&a_{l,+}^{+}|0\rangle_{A}a_{l,+}^{+}|0\rangle_{B}|\Phi^{+}\rangle_{12}|\Phi^{+}\rangle_{34}
\xrightarrow{M_{A}, Ions 1, 3, M_{B}, Ions 2, 4}\nonumber\\
&-\frac{1}{8}(a_{u,+}^{+}+ia_{l,+}^{+})|0\rangle_{A}(a_{u,+}^{+}+ia_{l,+}^{+})|0\rangle_{B}\nonumber\\
&\times|m_{+}\rangle_{1}|m_{+}\rangle_{2}|m_{-}\rangle_{3}|m_{-}\rangle_{4}\nonumber\\
&+\frac{1}{8}(a_{l,+}^{+}+ia_{u,+}^{+})|0\rangle_{A}(a_{l,+}^{+}+ia_{u,+}^{+})|0\rangle_{B}\nonumber\\
&\times|m_{-}\rangle_{1}|m_{-}\rangle_{2}|m_{+}\rangle_{3}|m_{+}\rangle_{4}\nonumber\\
&-\frac{1}{2}a_{u,+}^{+}|0\rangle_{A}a_{u,+}^{+}|0\rangle_{B}
|m_{-}\rangle_{1}|m_{-}\rangle_{2}|m_{-}\rangle_{3}|m_{-}\rangle_{4}\nonumber\\
&+\frac{\sqrt{10}}{4}|Scatter\rangle.
\end{eqnarray}
\begin{eqnarray}
&F(1-F):\nonumber\\
&a_{l,+}^{+}|0\rangle_{A}a_{l,+}^{+}|0\rangle_{B}|\Phi^{+}\rangle_{12}|\Psi^{+}\rangle_{34}
\xrightarrow{M_{A}, Ions 1, 3, M_{B}, Ions 2, 4}\nonumber\\
&\frac{i}{4}(a_{l,+}^{+}+ia_{u,+}^{+})|0\rangle_{A}a_{u,+}^{+}|0\rangle_{B}
|m_{-}\rangle_{1}|m_{-}\rangle_{2}|m_{+}\rangle_{3}|m_{-}\rangle_{4}\nonumber\\
&+\frac{i}{4}a_{u,+}^{+}|0\rangle_{A}(a_{l,+}^{+}+ia_{u,+}^{+})|0\rangle_{B}
|m_{-}\rangle_{1}|m_{-}\rangle_{2}|m_{-}\rangle_{3}|m_{+}\rangle_{4}\nonumber\\
&+\frac{\sqrt{3}}{2}|Scatter\rangle,
\end{eqnarray}
\begin{eqnarray}
&(1-F)F:\nonumber\\
&a_{l,+}^{+}|0\rangle_{A}a_{l,+}^{+}|0\rangle_{B}|\Psi^{+}\rangle_{12}|\Phi^{+}\rangle_{34}
\xrightarrow{M_{A}, Ions 1, 3, M_{B}, Ions 2, 4}\nonumber\\
&-\frac{1}{4}(a_{u,+}^{+}+ia_{l,+}^{+})|0\rangle_{A}a_{u,+}^{+}|0\rangle_{B}
|m_{+}\rangle_{1}|m_{-}\rangle_{2}|m_{-}\rangle_{3}|m_{-}\rangle_{4}\nonumber\\
&-\frac{1}{4}a_{u,+}^{+}|0\rangle_{A}(a_{u,+}^{+}+ia_{l,+}^{+})|0\rangle_{B}
|m_{-}\rangle_{1}|m_{+}\rangle_{2}|m_{-}\rangle_{3}|m_{-}\rangle_{4}\nonumber\\
&+\frac{\sqrt{3}}{2}|Scatter\rangle,
\end{eqnarray}
\begin{eqnarray}
&(1-F)^{2}:\nonumber\\
&a_{l,+}^{+}|0\rangle_{A}a_{l,+}^{+}|0\rangle_{B}|\Psi^{+}\rangle_{12}|\Psi^{+}\rangle_{34}
\xrightarrow{M_{A}, Ions 1, 3, M_{B}, Ions 2, 4}\nonumber\\
&\frac{i}{8}(a_{u,+}^{+}+ia_{l,+}^{+})|0\rangle_{A}(a_{l,+}^{+}+ia_{u,+}^{+})|0\rangle_{B}\nonumber\\
&\times|m_{+}\rangle_{1}|m_{-}\rangle_{2}|m_{-}\rangle_{3}|m_{+}\rangle_{4}\nonumber\\
&+\frac{i}{8}(a_{l,+}^{+}+ia_{u,+}^{+})|0\rangle_{A}(a_{u,+}^{+}+ia_{l,+}^{+})|0\rangle_{B}\nonumber\\
&\times|m_{-}\rangle_{1}|m_{+}\rangle_{2}|m_{+}\rangle_{3}|m_{-}\rangle_{4}
+\frac{\sqrt{14}}{4}|Scatter\rangle.
\end{eqnarray}
\end{subequations}
Where $|Scatter\rangle$ denotes the normalized vectors describing
the state of the scattered photons, which can be filtered out from
the detector. After evolution, Alice and Bob will operate single
photon measurements at the lower and upper output ports of $M_{A},
M_{B}$ respectively. In this purification scheme, the first
($|\Phi^{+}\rangle_{12}|\Phi^{+}\rangle_{34}$) and the fourth
($|\Psi^{+}\rangle_{12}|\Psi^{+}\rangle_{34}$) cases will lead to
the measurement result that the two lower output ports ($D_{Al}
\textrm{ and } D_{Bl}$) fire simultaneously, but the second
($|\Phi^{+}\rangle_{12}|\Psi^{+}\rangle_{34}$) and the third
($|\Psi^{+}\rangle_{12}|\Phi^{+}\rangle_{34}$)cases never lead to.
From the evolution result, we get that if the two lower output
ports ($D_{Al} \textrm{ and } D_{Bl}$) fire simultaneously, Alice
and Bob will get the four-ion maximally entangled state
$\frac{1}{\sqrt{2}}(|m_{+}\rangle_{1}|m_{+}\rangle_{2}|m_{-}\rangle_{3}|m_{-}\rangle_{4}
+|m_{-}\rangle_{1}|m_{-}\rangle_{2}|m_{+}\rangle_{3}|m_{+}\rangle_{4})$
with probability $\frac{F^{2}}{32}$, and get another four-ion
maximally entangled state
$\frac{1}{\sqrt{2}}(|m_{+}\rangle_{1}|m_{-}\rangle_{2}|m_{-}\rangle_{3}|m_{+}\rangle_{4}
+|m_{-}\rangle_{1}|m_{+}\rangle_{2}|m_{+}\rangle_{3}|m_{-}\rangle_{4})$
with probability $\frac{(1-F)^{2}}{32}$. If Alice and Bob measure
the ions 3 and 4 in the $|\pm\rangle$ basis, where
$|+\rangle=\frac{1}{\sqrt{2}}(|m_{+}\rangle+|m_{-}\rangle),
|-\rangle=\frac{1}{\sqrt{2}}(|m_{+}\rangle-|m_{-}\rangle)$, the
maximally entangled state:
$\frac{1}{\sqrt{2}}(|m_{+}\rangle_{1}|m_{+}\rangle_{2}|m_{-}\rangle_{3}|m_{-}\rangle_{4}
+|m_{-}\rangle_{1}|m_{-}\rangle_{2}|m_{+}\rangle_{3}|m_{+}\rangle_{4})$
will collapse into different states corresponding to various
measurement results. For the results
$|+\rangle_{3}|+\rangle_{4}\textrm{ and }
|-\rangle_{3}|-\rangle_{4}$, the four-ion maximally entangled
state will collapse into the state $|\Phi^{+}\rangle_{12}$. But
for the results$|+\rangle_{3}|-\rangle_{4}\textrm{ and }
|-\rangle_{3}|+\rangle_{4}$, it will collapse into
$|\Phi^{-}\rangle_{12}$, then Alice can operate a phase rotation
operation on ion 1 to convert $|\Phi^{-}\rangle_{12}$ into
$|\Phi^{+}\rangle_{12}$. For the four-ion maximally entangled
state
$\frac{1}{\sqrt{2}}(|m_{+}\rangle_{1}|m_{-}\rangle_{2}|m_{-}\rangle_{3}|m_{+}\rangle_{4}
+|m_{-}\rangle_{1}|m_{+}\rangle_{2}|m_{+}\rangle_{3}|m_{-}\rangle_{4})$
case, the measurement results and the needed operations have been
synchronized with the first case naturally. So after the evolution
, the single photon measurement and single ion measurement on each
side, the two remaining ions will be left in the new states
expressed by the new density operator:
\begin{equation}\label{new}
\rho_{12}=F^{'}|\Phi^{+}\rangle_{12}\langle\Phi^{+}|+(1-F^{'})|\Psi^{+}\rangle_{12}\langle\Psi^{+}|.
\end{equation}
where, $F^{'}=\frac{F^{2}}{F^{2}+(1-F)^{2}}$, is  the new
fidelity. If the fidelity of the initial shared entangled ensemble
satisfies $F>\frac{1}{2}$, $F^{'}>F$, the initial entangled state
is purified~\cite{jwp, jwp1}. Because $F$ can be an arbitrary
number between $0.5$ and $1.0$, the iteration of our scheme can
extract a near-perfect maximally entangled state from the ensemble
shared by Alice and Bob.

\section{ENTANGLEMENT CONCENTRATION FOR PURE NON-MAXIMALLY ENTANGLED STATES}
Here concludes the discussion of the entanglement purification for
mixed ionic states. We find that the above scheme can also be used
to concentrate the non-maximally entangled pure states. The setup
and the ionic level structure are all same to the mixed states
case. We can suppose the non-maximally entangled pure state is in
the form:
\begin{equation}
|\Psi\rangle_{AB}=
a|m_{+}\rangle_{A}|m_{-}\rangle_{B}+b|m_{-}\rangle_{A}|m_{+}\rangle_{B}.
\end{equation}
where $|a|^{2}+|b|^{2}=1$. Just like the mixed state case, two
pairs of ions($1, 2\textrm{ and }3, 4$) will be placed on $M_{A},
M_{B}$. The evolution of the total state of the system can be
expressed as:
\begin{eqnarray}
a_{l,+}^{+}&|0\rangle_{A}a_{l,+}^{+}|0\rangle_{B}|\Psi\rangle_{12}|\Psi\rangle_{34}
\xrightarrow{M_{A}, Ions 1, 3, M_{B}, Ions 2, 4}\nonumber\\
&\frac{iab}{4}(a_{u,+}^{+}+ia_{l,+}^{+})|0\rangle_{A}(a_{l,+}^{+}+ia_{u,+}^{+})|0\rangle_{B}\nonumber\\
&\times|m_{+}\rangle_{1}|m_{-}\rangle_{2}|m_{-}\rangle_{3}|m_{+}\rangle_{4}\nonumber\\
&+\frac{iab}{4}(a_{l,+}^{+}+ia_{u,+}^{+})|0\rangle_{A}(a_{u,+}^{+}+ia_{l,+}^{+})|0\rangle_{B}\nonumber\\
&\times|m_{-}\rangle_{1}|m_{+}\rangle_{2}|m_{+}\rangle_{3}|m_{-}\rangle_{4}\nonumber\\
&+\sqrt{\frac{2-|a|^{2}|b|^{2}}{2}}|Scatter\rangle.
\end{eqnarray}
After evolution, if the detectors $D_{Al}\textrm{ and }D_{Bl}$
fire, the four ions are left in maximally entangled state:
$\frac{1}{\sqrt{2}}(|m_{+}\rangle_{1}|m_{-}\rangle_{2}|m_{-}\rangle_{3}|m_{+}\rangle_{4}
+|m_{-}\rangle_{1}|m_{+}\rangle_{2}|m_{+}\rangle_{3}|m_{-}\rangle_{4})$
with probability $\frac{|a|^{2}|b|^{2}}{8}$. Although we probably
can get the four-ion maximally entangled states corresponding to
the measurement results: $D_{Au}\textrm{ and }D_{Bu}$,
$D_{Al}\textrm{ and }D_{Bu}$, $D_{Au}\textrm{ and }D_{Bl}$, we
should omit these results for the reason that the fire at the
upper outport($D_{Au}, D_{Bu}$) probably means the ions are not
precisely placed on the optical pathes.

If the initial non-maximally entangled state is in following form:
$a|m_{+}\rangle_{A}|m_{+}\rangle_{B}+b|m_{-}\rangle_{A}|m_{-}\rangle_{B}$,
the concentration will similarly succeed, provided that the
$D_{Al} \textrm{ and } D_{Bl}$ fire, and the successful
probability is also $\frac{|a|^{2}|b|^{2}}{8}$. After obtaining
the four-ion maximally entangled states, Alice and Bob can make
single ion measurement on ions 3, 4 in the basis $|\pm\rangle$
just like in the mixed states case. Then the remaining ions 1, 2
will be left in two-ion maximally entangled state. From analysis,
the successful probability for obtaining two-ion maximally
entangled state is still $\frac{|a|^{2}|b|^{2}}{8}$.

If we want to get four-ion maximally entangled states, there is no
need for us to operate the ionic measurement in the basis
$|\pm\rangle$. So our purification and concentration scheme can
not only generate two-ion maximally entangled states but also can
generate four-ion maximally entangled states. In this sense, the
present scheme is more efficient then the previous
scheme~\cite{jwp}. In Pan's scheme, the four-photon entangled
states can not be extracted, because the measurement on one pair
of photons are needed to complete the purification procedure,
otherwise we can not get to know whether the purification succeeds
or not. While, in our scheme, after the single photon measurement,
the purification process can concludes if we need four-ion
maximally entangled states. Then the four-ion maximally entangled
states can be used as a robust entanglement resource in Quantum
Communication. That is to say, our scheme is a purification scheme
without postselection measurement~\cite{wxb}.

\section{DISCUSSION}
\begin{figure}
\includegraphics[width=\columnwidth]{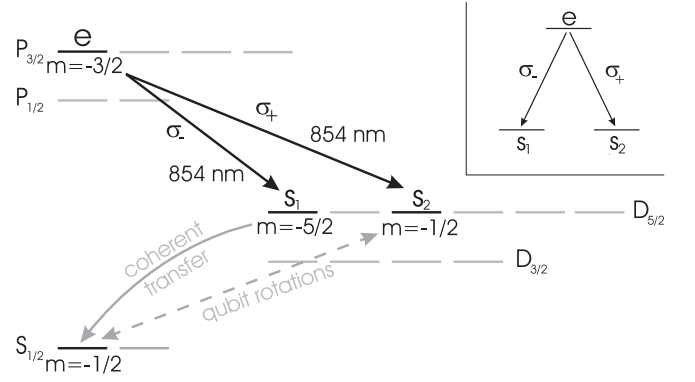}
\caption{Relevant levels of $^{40}$Ca$^{+}$ ions~\cite{ion3}.}
\label{fig3}
\end{figure}

After the discussion on the purification itself, we now discuss
the practical implementation of it. Singly positively charged
alkaline ions have only one electron outside a closed shell, so
they are commonly used in the quantum information experiments
using trapped ions~\cite{ion1,ion2}. Here we discuss a possible
implementation of our purification scheme using $^{40}$Ca$^{+}$ as
example. The relevant levels of $^{40}$Ca$^{+}$ has been Depicted
in Fig.\ref{fig3}~\cite{ion3}.

$D_{5/2}$ and $D_{3/2}$ are two metastable levels of
$^{40}$Ca$^{+}$ with lifetimes of the order of $1s$. $s_{1}$ and
$s_{2}$ are two sublevels of $D_{5/2}$ with $m=-5/2$ and $m=-1/2$,
and this two sublevels are coupled to $|e\rangle$ by $\sigma_-$
and $\sigma_+$ light at $854nm$. Here $e, S_{1}, S_{2}, S_{1/2}$
correspond to $e, m_{-}, m_{+}, g$ in Fig.\ref{level}
respectively. That is to say, we use the $S_{1/2}$ as stable
ground state, $S_{1}, S_{2}$ as two degenerate metastable state
and $P_{3/2}$ as excited state. Arbitrary superposition state of
this two degenerate metastable states can be realized by applying
a laser pulse of appropriate length, which can be realized in a
few microsecond~\cite{ion4}. The $^{40}$Ca$^{+}$ in state $S_{1}$
or $S_{2}$ can be excited into the excited state $P_{3/2}$ by
applying one $\sigma_-$ or $\sigma_+$ light at $854nm$. Then decay
from $|e\rangle$ to $S_{1}, S_{2}$, to $D_{3/2}$ and to $S_{1/2}$
are all possible. But References~\cite{ion2, ion3} give the
transition probability for $P_{1/2}\rightarrow S_{1/2}(397nm)$ as
$1.3 \times 10^8/$s and the branching ratio of $P_{1/2}\rightarrow
D_{3/2}(866nm)$ versus $P_{1/2}\rightarrow S_{1/2}(397nm)$ as
1:15, while the branching ratio for $P_{3/2}\rightarrow
D_{5/2}(854nm)$ versus $P_{3/2}\rightarrow S_{1/2}(393nm)$ can be
estimated as 1:30, giving $0.5 \times 10^7/$s for the transition
probability. So in most case, the $^{40}$Ca$^{+}$ in the excited
state will decay into the stable ground state $S_{1/2}$. The
detection of the internal states of $^{40}$Ca$^{+}$ can be
realized by using a cycling transition between $S_{1/2}$ and
$P_{1/2}(397nm)$~\cite{ion1,ion2}.

To enhance the emission efficiency of the photons from the ions,
we can introduce cavities. Then the following three items will
affect the emission efficiency of the photon from the ions:
\begin{itemize}
    \item The coupling between cavity mode and the $P_{3/2}\rightarrow
    S_{1/2}(393nm)$ transition;
    \item Decay from $P_{3/2}$ to $D_{5/2}$;
    \item Cavity decay.
\end{itemize}
From reference~\cite{decay}, the probability $p_{cav}$ for a
photon to be emitted into the cavity mode after excitation to $e$
can be expressed as $p_{cav}=\frac{4 \gamma
\Omega^2}{(\gamma+\Gamma)(\gamma \Gamma + 4 \Omega^2)}$. where
$\gamma=4\pi c/F_{cav}L$ is the decay rate of the cavity,
$F_{cav}$ its finesse, $L$ its length,
$\Omega=\frac{D}{\hbar}\sqrt{\frac{hc}{2 \epsilon_0 \lambda V}}$
is the coupling constant between the transition and the cavity
mode, $D$ the dipole element, $\lambda$ the wavelength of the
transition, $V$ the mode volume (which can be made as small as
$L^2\lambda/4$ for a confocal cavity with waist
$\sqrt{L\lambda/\pi}$), and $\Gamma$ is the non-cavity related
loss rate~\cite{ion3}. From the discussion of
reference~\cite{ion3}, the photon package is about $100ns$, which
is a relative long time for the purification scheme to be
completed.

When calculating the total efficiency of the purification scheme,
we must consider the following items:
\begin{itemize}
    \item The emission efficiency of photon: $p_{cav}$, which has included the cavity
    decay; To maximize the $p_{cav}$, we have chosen
    $F_{cav}=19000$, $L=3mm$. Then
    $\gamma=9.9\times10^{6}/s$, $p_{cav}=0.01$~\cite{ion3};
    \item The effect of the photon detectors is expressed as
    $\frac{{\eta}^{2}}{2}$.Here we let a detection efficiency
    $\eta=0.7$, which is a level that can be reached within the current
    technology.
    \item The asynchronism of the two users will also introduce
    some errors, which can be denoted by $\zeta$; The difficulty caused by the use of the two independent photon
sources can be solved by the following method. We can connect the
two users with a optical fibre. The photon source is placed in one
side, and the two twin photons produced by this source can be led
to two MZIs by this optical fibre. This method will also introduce
some error. Because the photon will transmit at the velocity of
light in the optical fibre, this error will be rather small. Then
the efficiency of the purification scheme will only be affected
slightly. So in numerical calculation, we can let $\zeta=0.9$. In
addition, we also can make use of a pair of classical pulses to
synchronize the two photon sources, where the pulses are used to
excite the optical switches.
    \item Coupling the photon out of the cavity will introduce
    another error expressed as $\xi$, which can be modulated to be
    close to unit.
\end{itemize}

After considering the above factors, the total success probability
can be expressed as follow:

\begin{itemize}
    \item
    $P={\frac{F^{2}+(1-F)^{2}}{32}}\times{p_{cav}}^{2}\times{\frac{{\eta}^{2}}{2}}\times{\zeta}\times{\xi}$
    for mixed state, that is to say, if we input photon with the
    rate of $3000000/s$, we can get seventy pairs of purified entangled $^{40}$Ca$^{+}$
    ions per minute for $F=0.7$.
    \item $P={\frac{a^{2}(1-a^{2})}{8}}\times{p_{cav}}^{2}\times{\frac{{\eta}^{2}}{2}}\times{\zeta}\times{\xi}$
    for pure state, that is to say, if we input photon with the
    rate of $3000000/s$, we can get one hundred pairs of purified entangled $^{40}$Ca$^{+}$
    ions per minute for $a^{2}=0.7$.
\end{itemize}

In conclusion, we present a entanglement purification scheme,
which can purify the general ionic mixed entangled stats by using
Mach-Zehnder interferometer. The most important advantage of the
scheme is that it can extract not only two-ion maximally entangled
states but also four-ion maximally entangled states from less
entangled pure or mixed states. In addition, the operations
carried out here are all simple and can be realized within the
current technology. The main drawback of the scheme is how to
place the ions on the optical pathes precisely.

\begin{acknowledgments}
This work is supported by Anhui Provincial Natural Science
Foundation under Grant No: 03042401, the Key Program of the
Education Department of Anhui Province under Grant No:2004kj005zd
and the Talent Foundation of Anhui University.
\end{acknowledgments}

\end{document}